\documentclass[a4paper,onecolumn,11pt,unpublished]{quantumarticle}
\pdfoutput=1
\usepackage[english]{babel}
\usepackage[T1]{fontenc}

\usepackage{tikz}
\usepackage{lipsum}

\usepackage{indentfirst}
\usepackage{mathtools}
\usepackage[numbers,sort&compress]{natbib}
\usepackage[utf8]{inputenc}
\usepackage{etoolbox}
\usepackage{color}
\usepackage{booktabs}

\usepackage{graphicx}
\usepackage{amssymb}   % for math
\usepackage{amsfonts}
\usepackage{amsmath}
\usepackage{dsfont}
\usepackage{natbib}
\usepackage{dcolumn}   % needed for some tables
\usepackage{bm}        % for math
\usepackage[mathscr]{eucal}
\usepackage{hyperref}

\makeatletter

\newcommand{\ep}{\varepsilon}

\newcommand{\bg}{{g}}
\newcommand{\bh}{{h}}

\newcommand{\be}{\begin{equation}}
\newcommand{\ee}{\end{equation}}
\def\ba{\begin{aligned}}
\def\ea{\end{aligned}}

\newcommand{\bea}{\begin{eqnarray}}
\newcommand{\eea}{\end{eqnarray}}

\newcommand{\bes}{\begin{subequations}}
\newcommand{\ees}{\end{subequations}}

\newcommand\mean[1]{\ensuremath{\left\langle#1\right\rangle}}
\newcommand\abs[1]{\ensuremath{\left|#1\right|}}

\newcommand\lrp[1]{\left(#1\right)}
\newcommand\lrb[1]{\left[#1\right]}

\newcommand\sgn[1]{\ensuremath{{\rm sign}\lrp{#1}}}
\newcommand{\lra}{\quad \Leftrightarrow \quad}

\renewcommand{\Re}{{\rm \, Re\,}}
\renewcommand{\Im}{{\rm \, Im\,}}

\renewcommand{\Im}{{\rm Im\,}}

\graphicspath{{./Figures/}}

\newcommand{\imk}[1]{{#1}}
\newcommand{\VM}[1]{{#1}}
\newcommand{\AG}[1]{{#1}}

\pdfoutput=1

\begin{document}
\title{Refined  cyclic renormalization group in  Russian Doll model}

\author{Vedant Motamarri}
\affiliation{TCM Group, Cavendish Laboratory, University of Cambridge, JJ Thomson Avenue, Cambridge, CB3 0HE, UK}
\orcid{0000-0001-9826-1177}
\author{Ivan M. Khaymovich}
\affiliation{Nordita, Stockholm University \& KTH Royal Institute of Technology, SE-106 91 Stockholm, Sweden}
\email{ivan.khaymovich@gmail.com}
\homepage{https://sites.google.com/view/ivan-khaymovich/}
\orcid{0000-0003-2160-5984}
\affiliation{Institute for Physics of Microstructures, Russian Academy of Sciences, 603950 Nizhny Novgorod, Russia}
\author{Alexander S. Gorsky}
\affiliation{Institute for Information Transmission Problems, Russian Academy of Sciences, 127051 Moscow, Russia}
\affiliation{Moscow Institute for Physics and Technology, Dolgoprudny 141700, Russia}
\affiliation{Laboratory of Complex Networks, Center for Neurophysics and Neuromorphyc Technologies, Moscow, Russia}
%\date{\today}%

\maketitle

\begin{abstract}
\imk{Focusing on Bethe-ansatz integrable models, robust to both time-reversal symmetry breaking and disorder, we consider the Russian Doll Model (RDM) for finite system sizes and energy levels.
\VM{Suggested} as a time-reversal-symmetry breaking deformation of Richardson's model, the well-known and simplest model of superconductivity, RDM revealed an unusual cyclic renormalization group (RG) over the system size $N$, where the energy levels repeat themselves, shifted by one after a finite period in $\ln N$, supplemented by a hierarchy of superconducting condensates, with the superconducting gaps following the so-called Efimov (exponential) scaling.
The equidistant single-particle spectrum of RDM made the above Efimov scaling and cyclic RG to be asymptotically exact in the wideband limit of the diagonal potential.
Here, we generalize this observation in various respects.
We find that, beyond the wideband limit, when the entire spectrum is considered, the periodicity of the spectrum is not constant, but appears to be energy-dependent.}
%\VM{Moreover, we resolve the paradox of the shift only by a single level after the RG period, where a finite fraction of energy levels disappears.}
%Moreover, we find the sink where the finite fraction of the energy levels disappears after a period of RG, whereas the superconducting condensates are shifted only by one.}
\VM{Moreover, we resolve the apparent paradox of shift in the spectrum by a single level after the RG period, despite the disappearance of a finite fraction of energy levels.}
%Moreover, we find that a finite fraction of energy levels disappears after a period of renormalization group (RG), while the superconducting condensates are only shifted by one
%VM: meaning not entirely clear.
\imk{We also analyze the effects of disorder in the diagonal potential on the above periodicity and show that it survives only for high energies beyond the energy interval of the disorder amplitude.
Our analytic analysis is supported with exact diagonalization.}
\end{abstract}
% TODO: include a table of contents (optional)
% Guideline: if your paper is longer that 6 pages, include a TOC
% To remove the TOC, simply cut the following block
% \vspace{10pt}
\noindent\rule{\textwidth}{1pt}
\tableofcontents\thispagestyle{fancy}
\noindent\rule{\textwidth}{1pt}
% \vspace{10pt}

\section{Introduction}
\imk{The Richardson model, suggested in~\cite{Richardson1963restricted,Richardson1964exact}
is a widely known and simple toy model of superconductivity with a fixed number of fermions/Cooper pairs in the condensate.
Besides the fact that it captures the main properties of superconductivity, this model is known to be integrable via the Bethe ansatz (BA). Indeed, the spectrum of the Richardson's model can be obtained via BA equations, which moreover, coincide with those for another well-known integrable model, namely, the twisted SU(2) Gaudin magnets~\cite{Cambiaggio1987integrability}.
%It is even known that the full set of commuting integrals of motion of the Richardson model get identified via superpositions of the Gaudin Hamiltonians.
Note that similar all-to-all Hamiltonians have been used in the physics literature for the description of superconducting grains~\cite{Matveev_Larkin1997} and the disorder-induced superconductor-insulator transition~\cite{Feigelman_Kravtsov_AnnPhys2010}.}

\imk{ As soon as any superconductor reveals itself not as a pure conductor, but as a pure diamagnet, the magnetic-field effects, such as the Meissner and Aharonov-Bohm ones, are most crucial manifestations of its superconducting properties.
In the all-to-all coupled (effectively zero-dimensional) Richarson's model, the effects of the magnetic field were investigated via a simple time-reversal (T) symmetry breaking deformation of the Richardson model~--~{the} so-called
Russian Doll model (RDM)~\cite{Leclair2004russian,Leclair2003russian}, where the all-to-all constant coupling $g$ gets the odd imaginary part $h$.
Already at that time, it appeared to be clear that RDM is also BA integrable, equivalent to the so-called twisted inhomogeneous XXX $SU(2)$ spin chain in terms of the BA equations
%The inhomogeneities get identified with the energy levels of RDM model while the twist is the counterpart of the coupling constant in RDM. The T-symmetry breaking parameter in RDM is identified as the "Planck constant" in XXX spin chain which vanishes in the Gaudin limit
\cite{Dunning2004integrability}, and can also be related to Chern-Simons theory when the excitations are represented by vertex operators~\cite{Asorey2002chern}.
}

\imk{
However, what was more surprising is that the RDM exhibits a rare property~---~it hosts a cyclic renormalization group (RG) flow
for the couplings via the system size $N$~\cite{Leclair2004russian}.
It was shown that in the wideband regime of the single-particle diagonal spectrum, the entire spectrum repeats itself over the finite period in $\ln N$.
Another peculiarity of RDM was that, unlike a single superconducting condensate in Richardson's model, it demonstrates an entire hierarchy of condensates with superconducting gaps, which are related to each other via a fixed exponential Efimov scaling and repeat (shifted by one) after the same RG period as the single-particle spectrum.
%The RG parameter is $\ln N$ where N is  number of energy levels and the step in RG is "integrating out" the highest energy level. The coupling evolves in a cyclic manner while the T-symmetry breaking parameter fixes the period of the cycle.
}

The above described RG cycle implies a nontrivial interplay between the ultraviolet- (UV) and infrared-limit (IR)  physics{,}
and the underlying algebraic property was identified as the anomalous
breaking of scale invariance down to the discrete subgroup~\cite{ananos2003anomalous,moroz2010nonrelativistic}.
It is this remaining discrete scale invariance in the models with
cyclic RG flows which is responsible for the fact that some part of the spectrum obeys the so-called Efimov exponential scaling $E_n\propto e^{cn}$, which, in the case of RDM, stands behind the hierarchy and periodicity of the superconducting gaps, see~\cite{Bulycheva2014spectrum} for a review.
Note that recently new examples of cyclic RG \cite{jepsen2021rg}{,} as well as
examples of homoclinic RG orbits \cite{jepsen2021homoclinic} and chaotic RG flows \cite{bosschaert2022chaotic} have been found.

\imk{
One of the questions which we address in this paper is related to the following paradox: how is it possible that the spectrum repeats itself (including the superconducting condensates) and shifts only by one level after the RG period $\Delta \ln N = \pi/h$, i.e., after the disappearance of the finite fraction of levels (but not just one)? Where do the other levels go?
}
\imk{
Another question is related to the robustness of both Richardson's model and RDM to diagonal disorder. Indeed, in both the models the BA equations are still applicable for any (even disordered) diagonal potentials.
However, this robustness of BA does not guarantee the corresponding robustness of the eigenstates or of the cyclic RG properties.
}

\imk{
Indeed, for the Richardson model with any strength of diagonal on-site disorder it is shown in~\cite{Ossipov2013anderson,Modak2016integrals,Nosov2019correlation,Nosov2019robustness,Kutlin2021emergent} to have all the excited eigenstates (except the superconducting ground state) to be power-law localized, while the eigenvalue statistics still indicates level repulsion.
%The measure for the ensemble averaging is a bit arbitrary and therelevance of the Generalized Gibbs Ensemble (GGE) for the disordered Richardson model has been discussed. The corresponding level statistics has been identified.
On the contrary, the corresponding eigenstates in RDM, considered in~\cite{motamarri2022localization},
show non-ergodic, but extended properties. Thus, the violation of time-reversal symmetry breaks the localization effects down and forms an entire fractal phase, similar to observations
in several other models~\cite{Kravtsov2015random,Biroli_RP,Ossipov_EPL2016_H+V,Monthus2017multifractality,vonSoosten2017non,Bogomolny2018eigenvalue,
Nosov2019correlation,Nosov2019robustness,Kutlin2021emergent,Tang2022nonergodic,%Kravtsov2020localization,Khaymovich2020fragile,Khaymovich2021dynamical,kutlin2023anatomy,
Biroli2021levy,motamarri2022localization,Buijsman2022circular}, with the Rosenzweig-Porter model (RPM) being the most familiar example.
}

\imk{
In this study, we address the same question of disorder effects in the context of the cyclicity of RG. We generalize the cyclic RG of the couplings, developed in~\cite{Leclair2004russian} for the case of equidistant spectrum, in two respects. First, we make a refinement of the cyclic RG, applicable for the entire spectrum beyond the wideband limit,  both for equidistant
and non-equidistant spectra, and find
that the cyclic RG structure survives but the period of the RG becomes energy-dependent.
Second, we incorporate the diagonal disorder into the derivation of the RG for the random RDM.
The analysis yields a similar result -- the period of the RG becomes energy-
and disorder-dependent and for the disorder potential, which reshuffles the order of the diagonal energies, the periodicity survives only in some parts of the spectrum.
We also comment on the fate of the Efimov tower and the incomplete breaking of scale invariance in these cases.
Thus, to summarize, the effects of disorder on the cyclic RG in RDM is not as straightforward as on the eigenstates or BA: the hierarchy of the condensates and, partially, the periodicity of RG survive, but the period depends both on the considered energy and disorder (and its concrete realization).
}

%Note that due to relation of the
%RDM with the inhomogeneous twisted XXX chain we could have in mind
%the disordered spin chain with ensemble of inhomogeneities. In the
%context of XXX spin chain the dependence on the inhomogeneities is
%governed by the KZ equations. On the other hand the inhomogeneous
%twisted XXX spin chain is related to the rational Ruijsenaars-Schneider(RS)
%model~\cite{Gorsky2012sqcd,Gorsky2014spectrum,Bulycheva2014spectrum} which provides the additional view on the ensemble
%averaging since inhomogeneities get identified with the coordinates
%in RS model.

Note that the T-breaking parameter {is usually not} renormalized
perturbatively, {but} can be renormalized however if some kind of
non-perturbative effects are taken into account. {An} example of RG {in a disordered} system with Anderson localization {and} T-breaking can be found in~\cite{Altland2015topology}.
%%%%%%%%%%%%%%%%%%%%%%%%%%%%%%%%%%%%%%%%%%%%%%%%%%%%%%%%%5
\section{Model}
%%%%%%%%%%%%%%%%%%%%%%%%%%%%%%%%%%%%%%%%%%%%%%%%%%%%%%%%%5

\imk{Here we introduce the Hamiltonian of the Russian Doll model.}
We consider the $N_0\times N_0$ random matrix model of the following form
\be\label{eq:model}
H_{mn} = \ep_n \delta_{mn} - j_{mn}, \quad
j_{m\ne n}=\delta(N_0) \lrb{g + i h \sgn{m-n}}, \quad 1\leq m,n\leq N_0 \ ,
\ee
\imk{where} we consider open boundary conditions and put the overall energy shift \VM{$j_{nn}$ to zero} without loss of generality.
Here $\ep_n$ is a certain (might be random and non-monotonic) potential of $n$ on a finite support
\be\label{eq:ep_n_support}
\abs{\ep_n} \leq \omega/2 \
\ee
\imk{and the matrix-size-dependent constant $\delta(N_0)$ is defined in the next section. }
%VM: this choice is different from Sierra, might lead to confusion in the next section

%%%%%%%%%%%%%%%%%%%%%%%%%%%%%%%%%%%%%%%%%%%%%%%%%%%%%%%%%5
\section{\imk{LeClair~-~Rom{\'a}n~-~}Sierra's Renormalization group (RG) for all energies.}
%%%%%%%%%%%%%%%%%%%%%%%%%%%%%%%%%%%%%%%%%%%%%%%%%%%%%%%%%5
\subsection{RG for equidistant spectrum}
In Ref.~\cite{Leclair2004russian} the authors consider the model~\eqref{eq:model} in the bosonic setting for application to superconductivity, with the following {choice for the parameter}
\be
\delta(N_0) = \omega/N_0 \ ,
\ee
with $\omega = \ep_{N_0} - \ep_1$ being the bandwidth~\footnote{Unlike~\cite{Leclair2004russian}, we use $\omega$ for the total bandwidth, not its half and $\delta$ for the level spacing, not its half.}
%VM: w is half the bandwidth in their model
%
of the diagonal potential. They focused on {the case with} equidistant spectrum of the diagonal potential
\be\label{eq:equidist}
\ep_n = (n-n_0) \delta \,
\ee
%VM: they use 2\delta as the gap between diagonal energies
with a certain energy shift $n_0 \delta$ (if not mentioned otherwise, we will use $n_0 = N_0/2$), giving the range of the diagonal energies as in~\eqref{eq:ep_n_support}.
They derived the following renormalization group (RG), see Eq.~(15) in Ref.~\cite{Leclair2004russian}, removing the largest diagonal energy {level} at each step:
\be\label{eq:Sierra_RG}
g_{N-1} = g_N +\frac{g_N^2+h_N^2}{N}, \quad {h_{N-1} = h_{N}} \ .
\ee
In order to derive the above equation{s} the authors of~\cite{Leclair2004russian} {did the following}:
\begin{enumerate}
  \item First, they start with the matrix of size $N_0$ and at each step reduce its size by one.
  \item For this, they take at each step the level with the {largest} diagonal energy in the absolute value ($\ep_N$ or $\ep_1$).
  \item Assuming it to be large with respect to the rest of the levels \imk{(a so-called wideband limit as we have mentioned it above)}, they resolve the eigenproblem with respect to it (say $\ep_N$):
\bes\label{eq:Large_E_RG}
\begin{align}
\lrp{\ep_N - E}\psi_E(N) - \sum_n j_{Nn} \psi_E(n) &= 0 \lra  \psi_E(N) = \frac{\sum_{n\ne N} j_{Nn} \psi_E(n)}{\ep_N - E}\\
\lrp{\ep_m - E}\psi_E(m) - \sum_n j_{mn} \psi_E(n) &= 0 \lra \nonumber\\
\lrp{\ep_m - E}\psi_E(m) &- \sum_{n\ne N} \lrp{j_{mn}+\frac{j_{mN} j_{Nn}}{\ep_N - E}} \psi_E(n) = 0
\label{eq:Large_E_RG_res}
\end{align}
\ees
Strictly speaking, the latter fraction
was split into two terms with $E$ replaced by $\ep_m$ and $\ep_n$, respectively, but this was not important for them.
  \item \label{item:bottom_E_assump} Next, they assumed $\ep_N - E \simeq \delta \cdot N$ and using the ratio $\omega/\delta = N$ they end up with Eq{s}.~\eqref{eq:Sierra_RG}.
\end{enumerate}
The solution of  Eqs.~\eqref{eq:Sierra_RG} can be found in the continuous limit {ds $\sim \Delta s =  - \Delta N / N\ll 1$}
\bes\label{period}
\begin{align}
h_{N}&= h_{N_0}\equiv h \ , \\
g_{N}&= h \tan\lrb{h s_N + \arctan\lrp{\frac{g_{N_0}}{h}}} \ .
\end{align}
\ees
with $\Delta N = 1$ and $s_N= \ln \lrp{N_0/N}$.
Strictly speaking the above RG works for the bottom of the spectrum $E\sim \ep_1$ if one takes the energies $\ep_N$ always from the top of the spectrum.

\imk{Physically, the solution~\eqref{period} means that the T-symmetry breaking parameter $h$ stays intact within such an RG over the logarithm of the system size $s_N = \ln \lrp{N_0/N}$, while the T-symmetric coupling $g_N$ changes periodically with the period $\Delta s = \pi/h$, determined by the T-breaking parameter $h$.}

\subsection{Energy dependent RG periods}\label{sec:general_res}
\imk{In order to go beyond applicability only for the bottom of the spectrum, mentioned in the end of the previous section and apply the results to the \emph{entire} spectrum, }
one should replace the assumption in  item~\ref{item:bottom_E_assump} by the correct \imk{energy-dependent} expression
\be\label{eq:RG-step}
g_{N-1} = g_N +\delta(N_0)\frac{g_N^2+h_N^2}{\ep_N - E}, \quad h_N = h_{N_0} \ .
\ee
Now the renormalization variable $s_E$ should be defined as
\be\label{eq:ds_res}
{d s_E(N)} = {-} \frac{\delta(N_0)}{\ep_N - E} \lra {s_E(N)} =
\sum_{n=N}^{N_0}\frac{\delta\cdot  \Delta N}{\ep_n - E} \approx
\int_N^{N_0} \frac{\delta \cdot dn}{\ep_n - E} \ ,
\ee
where $\Delta N = 1$ and one {arrives at} the same RG equations and solution as Eqs.(15-16) in~\cite{Leclair2004russian}
\be\label{eq:g(s)}
\frac{d\bg}{{ds_E}} = \bg^2+\bh^2 \lra
\boxed{
\bg({s_E}) = \bh \tan\lrb{\bh {{s_E}} + \arctan\lrp{\frac{\bg_{N_0}}{\bh}}}
} \ .
\ee
The validity of the above {equation}~\eqref{eq:g(s)} is limited by the condition{s} {for} the absence of resonance
\be\label{eq:ds<<1}
\abs{{{ds_E(N)}}}\ll 1, \; \frac{1}{g^2({s_E})+h^2} \lra \abs{E - \ep_N}\gg \delta, \;\delta\cdot\lrb{g^2({{s_E}})+h^2} \ .
\ee
The first condition ${\abs{d s_E(N)}}\ll 1$ ensures that the increment in the integral~\eqref{eq:ds_res} is small,
while the second one limits the increment
%VM: general query: do we really need the second condition i.e. |dg(s)| <<1? Even for Sierra's RG, i.e. for E close to 0, we see that dg(s)-->\inf is possible whenever h*s + arctan(g_0/h) --> \pi.
% Essentially g(s) blows up and so does |dg(s)| during the RG.
$\abs{dg(s)}$ in~\eqref{eq:g(s)} to make the derivation from~\eqref{eq:RG-step} to it valid.
Note that from Eq.~\eqref{eq:ds_res} one can see that the monotonicity of the parameter $s$~\eqref{eq:ds_res} depends on the energy $E$ and does not necessarily {require} the monotonicity of $\ep_n$.
Indeed, for $|E|>\omega/2>|\ep_n|$, even random $\ep_n$ does not change the monotonic behavior of $s_E(N)$\imk{, keeping the periodicity of the cyclic RG robust in this energy interval}. In the following two subsections, we apply the above considerations for equidistant and disordered diagonal potentials.

\subsection{Entire spectrum for equidistant potential}
For equidistant diagonal potential~\eqref{eq:equidist}, one can introduce the following {parameter} $M_E = E/\delta + n_0$, {for} the energy shift, which gives:
\be\label{eq:sN_equidist}
{s_E(N)} = \ln\lrp{\frac{N_0 - M_E}{N - M_E}}.
\ee
As in Eq.~(17) of~\cite{Leclair2004russian} the result~\eqref{eq:g(s)} is periodic with the period $\lambda = \pi/\bh$ in ${s_E}$.
The period $\lambda$ in ${s_E}$ corresponds to the change $\Delta N_T$ in the matrix size $N$ given by
\be\label{eq:dN}
\lambda = \ln\lrp{\frac{N-M_E}{N-\Delta N_T - M_E}} \ , \quad
\boxed{\Delta N_T =\lrp{N-M_E}\lrp{1-e^{-\lambda}}.}
\ee
The number of periods before $N-M_E=1$ goes as
\be\label{eq:Sierra_n_condensates}
\boxed{n = \frac{h}{\pi} \ln\lrp{N_0 - M_E}.}
\ee
%Note that any $N_0$-dependence of $\delta$ just rescales the period $\lambda$, but does not change the logarithmic in $N$ periodicity.
However, unlike~\cite{Leclair2004russian}, here we see two peculiarities:
\begin{itemize}
  \item First, the period $\Delta N_T$ in $N$ is energy $E$-dependent, Eq.~\eqref{eq:dN}, and
  \item Second, there is the singularity at $N=M_E$, or equivalently, at $E=\ep_N$.
\end{itemize}

The latter is important for the understanding of the \imk{first question (or paradox), mentioned in the introduction}.
Indeed, according to~\cite{Leclair2004russian} and Eq.~\eqref{eq:dN}, the matrix size shrinks by $\Delta N_T(N)$ \imk{after the period, when} the spectrum repeats itself with the shift by one level.
However, on the way from $N_0$ to $N$ other $\Delta N_T-1$ levels have also disappeared. Where have they gone?

To answer this physical question, one should consider the continuity condition~\eqref{eq:ds<<1} more {closely}. What happens when this condition is violated? In such a case, one cannot transform the sum in Eq.~\eqref{eq:ds_res} to the integral and, moreover, already at \imk{a single} step, one of the increments $ds_E(N)$ or $dg({s_E})$ is not small.
This means that many periods can pass in this region without being seen in \imk{the continuous equations~\eqref{eq:g(s)} and the} numerics.
Strictly speaking, each RG step~\eqref{eq:RG-step} corresponds to the removal of one column and one row of the matrix and can be considered as a rank-$1$ perturbation for the matrix~\cite{Bogomolny_rank1}. As it is known from Richardson's model~\cite{Modak2016integrals} and other works~\cite{Kutlin2021emergent,Kochergin2024robust}, such a rank-$1$ perturbation can move significantly only one (top or bottom) level $E_N^{(N-1)}$, while the other $N-2$ levels $E_n^{(N-1)}$ are bound in between the ones at the previous step
\be
E_n^{(N)}<E_n^{(N-1)}<E_{n+1}^{(N)} \ .
\ee
This means that independently of the condition~\eqref{eq:ds<<1} only one level disappears from the spectrum by going to the \imk{sink} at $E=\ep_N$ in the RG step.
We will show the same in our numerical results in {Sec.~\ref{sec:numerics}}.

\subsection{Case of the diagonal disorder}
Strictly speaking Eqs.~\eqref{eq:RG-step} and~\eqref{eq:ds_res} work for any diagonal potential, not only for {e}quidistant or monotonic $\ep_n$. {T}herefore in this subsection we consider disordered diagonal potential. In the case when the diagonal energies are not equidistant~\eqref{eq:equidist} but given by independent random numbers,
the derivation of RG equations from~\eqref{eq:Large_E_RG} is not completely trivial.

In order to make the derivation clear let's consider separately the two effects of the {disorder}:
%VM: removed the's from both items below
\begin{enumerate}
    \item Fluctuations of $\ep_n$ around their mean value~\eqref{eq:equidist};
    \item Re-shuffling of $\ep_n$.
\end{enumerate}
Taking into account only the first effect\imk{, i.e., keeping the order $\ep_n\leq\ep_{n+1}$}, one can represent
$\ep_n$ as a sum of independent \imk{non-negative} increments
\be
\ep_n = \ep_1+\sum_{k=1}^{n-1} \delta \ep_k \ , \quad
P(\{\delta \ep_k\}) = \prod_{k=1}^{N-1} P_0(\delta \ep_k) \ , \quad
P_0(x) = \frac{1}{\delta} e^{-x/\delta} \ , \quad \mean{\delta \ep_k} = \delta \ .
\ee
For large enough $n_E = n-n_0-E/\delta\equiv n - M_E\gg 1$, Eq.~\eqref{eq:equidist},
of i.i.d. random elements in the sum $\ep_n - E$, {can be approximated} by a Gaussian random number with the following mean and variance
\be\label{eq:en_random}
\mean{\ep_n  - E} = \delta \cdot n_E \ , \quad \sigma^2_{n,E} = \mean{(\ep_n - E)^2} - \mean{\ep_n - E}^2 = \delta^2 \cdot n_E \ ,
\ee
and thus can be represented as
\be\label{eq:epN_CLT}
\ep_n= E + \delta \cdot n_E + \delta \cdot \sqrt{n_E} G_n = \delta\lrp{n - n_0} + \delta \cdot \sqrt{n_E} G_n \ ,
\ee
with the standard Gaussian variable $G_n$
\be\label{eq:Gauss_std}
\mean{G_n} = 0 \ , \quad \mean{G_n^2}=1 \ .
\ee

The corresponding increment ${ds_E(N)}$, Eq.~\eqref{eq:ds_res}, is then given by %(with $\Delta N = 1$)
\be\label{eq:ds_N_random}
{ds_E} = \frac{\Delta N}{n_E\lrp{1 + \imk{G_N/\sqrt{n_E}}}} \simeq
\frac{\Delta N}{n_E}\lrp{1 - \frac{G_N}{n_E^{1/2}}}
\ee
With the latter Taylor expansion this gives the result for {$s_E(N)$} in terms of the central limit theorem as
\begin{multline}\label{eq:sN_disorder_monotonic}
{s_E(N)} = \ln\lrp{\frac{N_0 - M_E}{N - M_E}} - \tilde G_N \lrp{\sum_{n=N}^{N_0}\frac{1}{\lrp{n-M_E}^3}}^{1/2} \\\simeq
\ln\lrp{\frac{N_0 - M_E}{N - M_E}} - \frac{\tilde G_N}{\sqrt{2}\abs{N-M_E}}\lrb{1-\lrp{\frac{N-M_E}{N_0-M_E}}^2}^{1/2} \ ,
\end{multline}
Here we used the central limit theorem for the sum of Gaussians $G_n/n_E^{3/2}$ with zero means and variances $\sigma_n^2 = n_E^{-3}$ and introduced another standard Gaussian variable $\tilde G_N$, Eq.~\eqref{eq:Gauss_std}.

From the latter one can see that the additional summand $\sim |N-M_E|^{-1}$ to $s_N$ with respect to the one in the disorder-free case, Eq.~\eqref{eq:sN_equidist}, is small compared to the period $\pi/h$ for large enough $N-M_E$ within the RG validity region, Eq.~\eqref{eq:ds<<1}.
Strictly speaking, in the sum~\eqref{eq:sN_disorder_monotonic} one cannot keep the terms {$O(1)$} as the Euler-Mascheroni constant $\gamma_E \simeq 0.5772$ is also neglected there.

At the same time, at the top of the spectrum (from where we take out $\ep_N$) and close to \imk{$N\approx M_E$, }
the fluctuations will be important already at the level of Eq.~\eqref{eq:ds_N_random}.
In the former region the central limit theorem in~\eqref{eq:epN_CLT} does not {hold},
while in the latter the entire validity of the RG~\eqref{eq:ds<<1} is broken. As a result, \imk{with this we show that the periodicity of RG for} the spectrum survives in the monotonic but disordered diagonal potential \imk{within the same validity range away from the sink point $E = \ep_N$, i.e., at $\abs{E - \ep_N}\gg E$}.

The reshuffling of the diagonal disorder has another effect.
Indeed, as we mentioned in Sec.~\ref{sec:general_res}, in this case, the monotonicity of the periodicity parameter ${s_E(N)}$ is guaranteed only for $|E|>\omega/2$. \imk{Otherwise, both the sign and the amplitude of the increment $ds_E(N$ in Eq.~\eqref{eq:ds_res} are random and the validity conditions~\eqref{eq:ds<<1} to derive the continuous equation~\eqref{eq:g(s)} cannot be satisfied.}
Therefore, the above periodicity~\eqref{eq:g(s)} survives only in \imk{the above mentioned region $|E|>\omega/2$}, while within the diagonal band, $|E|<\omega/2$, the parameter {$s_E(N)$} can be non-monotonic with $N$ and random, and therefore no periodicity is expected.

\imk{To sum up this section, we showed that the effect of the diagonal potential fluctuations without reshuffling affects only the vicinity of the sink point and the cyclic RG survives in the same validity range as for the equidistant spectrum. At the same time, reshuffling the diagonal elements ruins the periodicity of the RG in the entire range of the diagonal potential $|E|<\omega/2$, keeping it intact only beyond it, including the condensate energies (and the corresponding Efimov scaling, as we will see below).}

\subsection{Generalized Efimov scaling}

Let us comment on the place of our study in the general context of the
two-parametric RG flows when one parameter induces T-symmetry breaking. It is useful
to introduce the following modular parameter \cite{flack2023generalized}
\be
\tau =x+iy \text{, where $x$ is T-symmetry breaking term and $y$ is a some kind of disorder}
\ee
The real part is the chemical potential for the topological number of any
nature, say, winding, topological charge etc. On the other hand the imaginary part
is any parameter quantifying disorder, say, coupling constant, diffusion coefficient,
boundary condition etc. In our case one could have in mind $\tau= h + i g$
while, for example , $\tau= \theta+iD$ for the Anderson model in 1d with T-symmetry breaking
term $\theta$ and the diffusion coefficient $D$ \cite{Altland2015topology}. Before renormalization
there is the natural action of $SL(2,Z)$ on modular domain of $\tau$.

The pattern of RG orbits considered as trajectories of the dynamical systems
depends on the relative weights of the perturbative and non-perturbative
contributions to the $\beta$-functions.  The
conventional cyclic RG occurs at stable fixed point for $\Re\tau$  taking
into account only a
first perturbative contribution to the  $\beta_{\Im\tau}$.
The $\Re\tau$ is finite at the stable fixed point  and it governs the period of the RG cycle~\eqref{eq:g(s)}.
Generically both the $\beta$'s are elliptic functions of the
modular parameter and behave differently in the limiting cases.

The RG flow toward{s} the stable fixed point can occur through the chain
of unstable fixed points for $\Re\tau$.
For instance, such behavior and interesting universality has been observed in \cite{flack2023generalized}
in the limit $y=\Im \tau \rightarrow 0$ when the potential function for the RG flow which yields the $\beta(x,y)$ function
is the generalized Dedekind function.
\be
U(x,y)= \log(y^{1/4} |\eta(x+iy)|                                 \qquad \tau=x+iy \ ,
\ee
where $\eta(z)$ is Dedekind function $\eta(z)= e^{\frac{\pi i z}{12}}\prod_{n=0}^{\infty}( 1- e^{2\pi i nz})$.
At small fixed $y$ the RG potential for T-breaking parameter gets reduced to $U(x)$ whose
minima $x_n$ exhibit the interesting recurrence $x_{n+1}= f(x_n)$ for the unstable critical
points of the RG flow for $\Im\tau$. The recurrence is ruled by
the free group $\Gamma_2$ which is subgroup of $SL(2,Z)$ and involves three
generators of the discrete RG flows \cite{flack2023generalized}.
At $n\rightarrow  \infty$ the RG {flows} to the stable critical point
while a topological parameter
tends to the Golden ratio $x_n\rightarrow \frac{\sqrt{5}+1}{2}$.
This model example corresponds to
the one-dimensional Penrose model which is a toy model
{exhibiting} cyclic RG cycle.
In that case the Efimov scaling for the bound states reads as
\be
E_n=E_0\exp(cn)
\label{efimov}
\ee
with $c= \ln(\frac{\sqrt{5}+1}{2})$. In the refined RG, {o}nce again we look at the stable
fixed point of $\Re\tau$ but the period of $\Im\tau $
is energy dependent
\be
g(s_E + \lambda)=g(s_E) \ .
\ee
A bit loosely we could say that the
$\Re\tau$ defining the period at the fixed point is E-dependent.
Instead of (\ref{efimov}) we have scaling of the form
\be
\frac{\log(\frac{E_n}{E_0})}{s(E_n)}\sim n
\ee
which reduces to Efimov scaling for equidistant spectrum.
{Note} that the Efimov scaling follows from the partial
breaking of the scale invariance down to the
discrete
subgroup \cite{ananos2003anomalous,moroz2010nonrelativistic}. In the refined case the discrete subgroup is
broken as well.

\AG{It is worth making one more remark. In~\cite{Leclair2003russian}, the set of resonances in the particular $(1+1)$ quantum field theory (QFT) with the scaling
\be
    M_n=2m\cosh(\lambda n)
\ee
where the $\lambda = \pi/h$ is a period of the peculiar RG flow, has been found.
Certainly it does not enjoy the Efimov scaling at low energy but the spectrum can  be represented in the form of the generalised Efimov scaling with the energy dependent RG period $\lambda(n)$
\be
    M_n=2m\exp(\lambda(n) n), \qquad \lambda(n)= \lambda+ \frac{1}{n}
    \log\frac{(1+ e^{-2n\lambda})}{2}
\ee
Hence it seems that our finding in finite dimensional system has the clear QFT counterpart however it would be nice to investigate this relation in more details.}

\section{Numerics}
\label{sec:numerics}
\begin{figure}[h!]
    \includegraphics[width=\columnwidth]{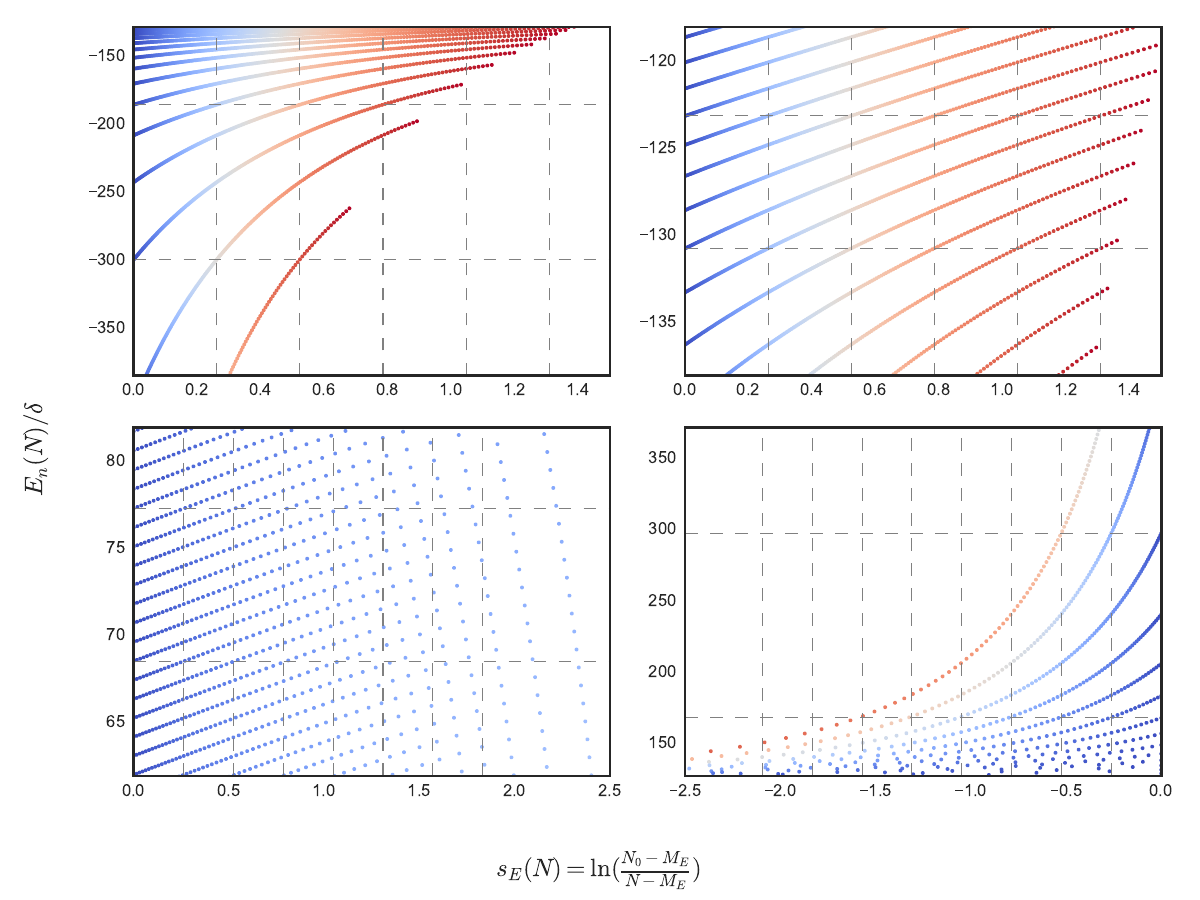}
    \caption{\textbf{Generalized $E$-dependent spectrum periodicity, Eq.~\eqref{eq:sN_equidist}, in the Russian Doll model with equidistant diagonal potential, Eq.~\eqref{eq:equidist}} in different spectral parts.
    Vertical lines correspond to the periodicity in the parameter ${s_E(N)} = \ln\lrp{\frac{N_0 - M_E}{N - M_E}}$, which perfectly matches the one in the numerical spectrum in all those parts. \VM{The color of the data points varies from blue to red as the system size is reduced from $N_0=256$ to 64.}
    }
    \label{fig:equidist}
\end{figure}

\begin{figure}[h!]
\includegraphics[width=\columnwidth]{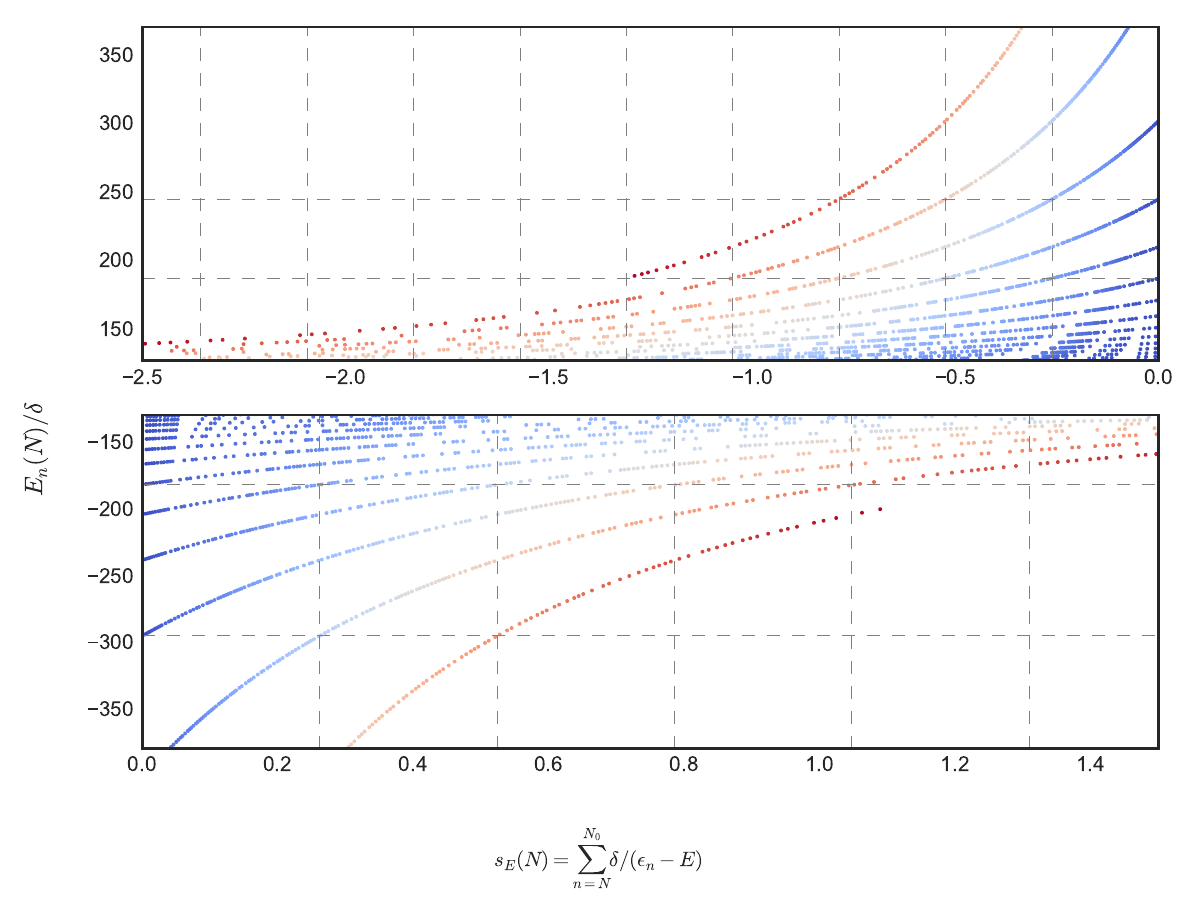}
\caption{\textbf{Generalized $E$-dependent spectrum periodicity, Eq.~\eqref{eq:en_random}, in the Russian Doll model with random diagonal potential} in different spectral parts.
Vertical lines show the periodicity in the parameter $s_E(N)$~\eqref{eq:ds_res}, which provides reasonable match to the periodicity of the spectrum in the parts, away from the diagonal potential bulk, $|E|>\omega/2$. %Here the diagonal potential does not have the shift {$n_0 = 0$} and distributed in the interval {$0<\ep_n<\omega$}.
\VM{The color of the data points varies from blue to red as the system size is reduced from $N_0=256$ to 64.}
}
\label{fig:disorder_parts}
\end{figure}

\begin{figure}[h!]    \includegraphics[width=0.95\columnwidth]{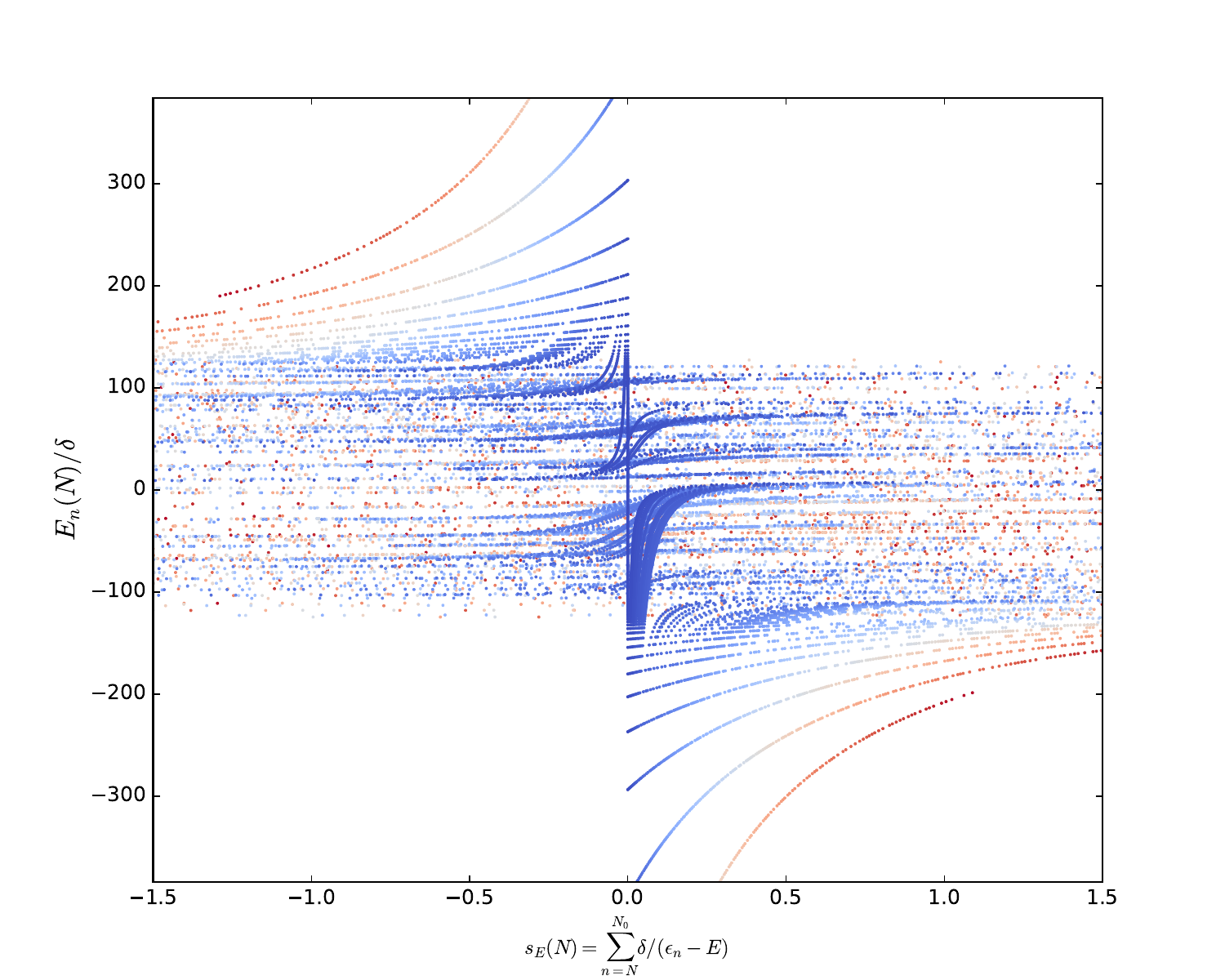}
    \caption{\textbf{Overview of generalized $E$-dependent spectrum periodicity in the Russian Doll model with random diagonal potential.}
    One can clearly see that the periodicity is broken within the diagonal potential bulk, $|E|<\omega/2$.
    % Here the diagonal potential {does not} have the shift {$n_0 = 0$}.
    \VM{The color of the data points varies from blue to red as the system size is reduced from $N_0=256$ to 64.}
    }
    \label{fig:disorder_overview}
\end{figure}

In order to check the analytical predictions of the previous sections, we have performed numerical simulations similar to~\cite{Leclair2004russian}. Taking the initial Hamiltonian~\eqref{eq:model} of size $N_0$, we compute the spectrum for the models, given by the first $N$ rows and $N$ columns of the matrix.
Then the spectrum of such models (normalized by the parameter $\delta$) has been plotted versus the periodicity parameter ${s_E(N)}$, Eq.~\eqref{eq:ds_res}, see Figs.~\ref{fig:equidist}~--~\ref{fig:disorder_overview}.
For our numerics we have chosen $N_0 = 256$, $g=1$, and $h=12$, though the results are qualitatively the same for other parameters as well.
%We also consider {$n_0 = 0$ to compare them} with~\cite{Leclair2004russian}.

In the case of the equidistant spectrum, Fig.~\ref{fig:equidist}, the periodicity parameter is given by~\eqref{eq:sN_equidist},  $s_E(N) = \ln\lrp{\frac{N_0 - M_E}{N - M_E}}$, with $M_E = E/\delta + n_0$.
One can see from Fig.~\ref{fig:equidist} that the \imk{period varies} in different regions of the spectrum, but it is still given by the above formula.
At the spectral edges (see the first and last panels), the energy levels may disappear with decreasing $N$ (and increasing $s_E(N)$).
In addition, close to the energies $E/\delta = N \equiv \ep_N/\delta$, (see the bottom part of the last panel in Fig.~\ref{fig:equidist}), the periodicity is violated in full agreement with the validity range, Eq.~\eqref{eq:ds<<1}.
Note that it is not just the {discreteness of the spectrum which matters as for smaller energies $|E|/\delta < N$}
%VM: does this statement need to be changed since the right bottom panel is not for E/\delta < N ?
even the discrete spectrum shows the same periodicity, cf. left ($|E|/\delta < N$) and right ($|E|/\delta > N$) bottom panels.

In the more physical and interesting case of disordered diagonal potential~\eqref{eq:en_random}, one has to modify the periodicity parameter to~\eqref{eq:ds_res} or in the monotonic case to~\eqref{eq:sN_disorder_monotonic}.
In this case, see Fig.~\ref{fig:disorder_parts}, the periodicity is still clearly seen, but close to the interval of the diagonal potential energies, $|E|<\omega/2$ (with $n_0=N_0/2$) the periodic levels are not seen under the ones with random shifts along ${s_E(N)}$. The latter are those levels, which hit the resonance $E \simeq \ep_N$ and, thus, have non-monotonic $s_E(N)$ vs $N$.
%{They appear on top of the spectral edge features (see all panels, except the first one)}, where the levels may disappear with decreasing $N$ (increasing $s_E(N)$).

In order to show clearly the range of random energies, we plot the entire spectrum of the system in Fig.~\ref{fig:disorder_overview} versus the local periodicity parameter ${s_E(N)} = \sum_{n=N}^{N_0} \delta/(\ep_n -E)$. From that figure the periodicity is hard to see due to the symbol sizes, but one can clearly observe that in the interval $|E|/\delta<N_0/2$ the random levels, corresponding to the above hitting of resonances and non-monotonic $s_E(N)$, prevails over the the regular ones, so the latter are not seen.
Beyond the above mentioned energy interval, i.e. for $|E|/\delta>N_0/2$, no such random levels are visible \imk{and the regular (periodic) behavior is present}.

In addition, a random singular point of the spectrum appears at $E = \ep_N$, where the regular spectral part changes behavior from $s_N>0$ at $E<\ep_N$ to $s_N<0$ otherwise. These are exactly the sink points which are random within the interval $|\ep_N|<\omega/2$ at each step $N$ where most of the levels disappear beyond RG periodicity.

%%%%%%%%%%%%%%%%%%%%%%%%%%%
\section{Conclusion}
%%%%%%%%%%%%%%%%%%%%%%%%%%%
In this paper we have generalized the periodic renormalization group (RG) for the known Russian Doll model (RDM) in several \imk{respects}.

\imk{Within the original RDM setting} with the equidistant diagonal elements, we have shown that the RG period depends significantly on the energy interval \VM{considered} and has a singularity at the sink point $E = \ep_N$. It is this singularity which compensates the disbalance of $\Delta N_T - 1$ energy levels that should disappear after the RG period $\Delta N_T \simeq$ and, according to the previous literature~\cite{Leclair2004russian,Leclair2003russian}, shift the entire spectrum by one level only.

In addition, we have considered the RDM with the disordered diagonal elements and found \imk{two separate effects of disorder. First, the fluctuations of diagonal potential do not affect the validity of the cyclic RG, while the second reshuffling contribution of the diagonal potential allows the RG periodicity to survive only beyond the diagonal disorder amplitude.
For this, we derive }
a generalized RG parameter over which RG equations are still periodic  (at least in the spectral parts lying beyond the energy interval of the diagonal elements).
%We have considered both the effects of small oscillations of the diagonal elements as well as their re-shuffling due to disorder.
All the analytical predictions have been confirmed by the numerical simulations.

\imk{In the further investigations, it} would be interesting to identify the limit cycle breaking discussed in this study with the generic framework of breakdown of the limit cycle within the bifurcation theory.

\imk{The effect of periodicity, suggested in the Russian Doll model and generically considered in this work, is in some sense similar to the Aharonov-Bohm effect in the single-mode superconducting ring pierced by a magnetic flux $\phi$, where the (all-to-all site) coupling also has the real $g$ and imaginary $h$ parts, periodically changing with $\phi$, however, unlike the latter, the cyclic RG in RDM shows periodicity over the logarithm $\ln N$ of the system size and this periodicity, as we have shown, is energy-dependent.
In this sense, it will be of particular interest to find similar periodicity effects in some physical short-range models.
Among possible candidates one can guess to have hierarchical structures like the so-called random-regular or Erd\"os-Renyi graphs, where the dominant cycle size where the magnetic field can penetrate is, indeed, proportional to $\ln N$ and in this respect one can expect it to show similar Aharonov-Bohm periodicity as the magnetic flux will contain the dominant loop size $\sim \ln N$.
}
%%% OUTLOOK!!!

%%%%%%%%%%%%%%%%%%%%%%%%%%%
\section{Acknowledgments}
%%%%%%%%%%%%%%%%%%%%%%%%%%%
V.~M. thanks MPIPKS for providing an opportunity \VM{to participate} in the Summer Internship program in 2022 during which this project was initiated and acknowledges support from the Harding Foundation. I.~M.~K. acknowledges support by the Russian Science Foundation (Grant No. 21-72-10161). A.~S.~G. thanks Nordita and IHES where the parts of this work have been done for the hospitality and support.
\clearpage

\appendix

%%%%%%%%%%%%%%%%%%%%%%%%%%%%%%%%%%%%%%%%%%%%%%%%%%%%%%%%%5

\bibliography{RDM}
%%%%%%%%%%%%%%%%%%%%%%%%%%%%%%%%%%%%%%%%%%%%%%%%%%%%%%%%%5

\end{document}